\documentclass[conference]{IEEEtran}
\IEEEoverridecommandlockouts
\usepackage{cite}
\usepackage{amsmath,amssymb,amsfonts}
\usepackage{algorithm,algorithmic}
\usepackage{multirow}
\usepackage{graphicx}
\usepackage{textcomp}
\usepackage{xcolor}
\usepackage{flushend}
\def\BibTeX{{\rm B\kern-.05em{\sc i\kern-.025em b}\kern-.08em
    T\kern-.1667em\lower.7ex\hbox{E}\kern-.125emX}}
\begin{document}

\title{General Framework for Reversible Data Hiding in Texts Based on Masked Language Modeling}
\author{Xiaoyan Zheng$^2$, Yurun Fang$^2$ and Hanzhou Wu$^{1,2,*}$\\
$^1$Guangdong Provincial Key Laboratory of Information Security Technology, Guangzhou 510006, China\\
$^2$Shanghai University, Shanghai 200444, China\\
$^*$h.wu.phd@ieee.org\\
}

\maketitle

\begin{abstract}
With the fast development of natural language processing, recent advances in information hiding focus on covertly embedding secret information into texts. These algorithms either modify a given cover text or directly generate a text containing secret information, which, however, are not reversible, meaning that the original text not carrying secret information cannot be perfectly recovered unless much side information are shared in advance. To tackle with this problem, in this paper, we propose a general framework to embed secret information into a given cover text, for which the embedded information and the original cover text can be perfectly retrieved from the marked text. The main idea of the proposed method is to use a masked language model to generate such a marked text that the cover text can be reconstructed by collecting the words of some positions and the words of the other positions can be processed to extract the secret information. Our results show that the original cover text and the secret information can be successfully embedded and extracted. Meanwhile, the marked text carrying secret information has good fluency and semantic quality, indicating that the proposed method has satisfactory security, which has been verified by experimental results. Furthermore, there is no need for the data hider and data receiver to share the language model, which significantly reduces the side information and thus has good potential in applications. 
\end{abstract}


\section{Introduction}
As an effective means to secret communication, \emph{information hiding} (typically also called \emph{data hiding}) enables us to covertly embed a secret message in a digital media by taking advantage of the redundancy of the digital media. The newly generated media containing secret information will not introduce noticeable artifacts, resulting in that the usage of the media will not be impaired and additional purposes such as secret information transmission and copyright protection can be achieved. Though the secret information can be successfully extracted for many data hiding methods, permanent distortion will appear in the altered media content, which indicates that the original media content cannot be perfectly recovered which is not desirable for sensitive applications that require the original media to be fully recovered at the receiver side.

\emph{Reversible data hiding (RDH)} \cite{RDH:TCSVT:2017} is therefore deeply studied to deal with the above problem in recent years. Compared with traditional data hiding (DH) algorithms, RDH allows the data receiver to restore the original media content without distortion after extracting the embedded information from the embedded media. That is to say, after secret transmission via a lossless channel, when the user is authorized to obtain the same key, the original cover and the secret information can be recovered losslessly from the media containing the secret information. At present, RDH has been widely applied to digital images and video sequences \cite{RDH:PPE:2016, RDH:DCSPF:2016, MTAP:RDH:Video, Chen:IJDSN}, but there are very few achievements on RDH in texts. The reason is that text has the highly coded nature with little redundancy, and there are rather fewer bits of information available for embedding than images and videos. Therefore, it is necessary to design novel schemes for RDH in texts. Among many media types, text is the most commonly used information carrier by human beings. Moreover, text also has the property of high robustness with little change through channel transmission, and is not easily disturbed by noise. It means that RDH in texts has good application prospect, which motivates the authors in this paper to study RDH in texts. 

\begin{figure}[!t]
\begin{center}
\includegraphics[width=\linewidth]{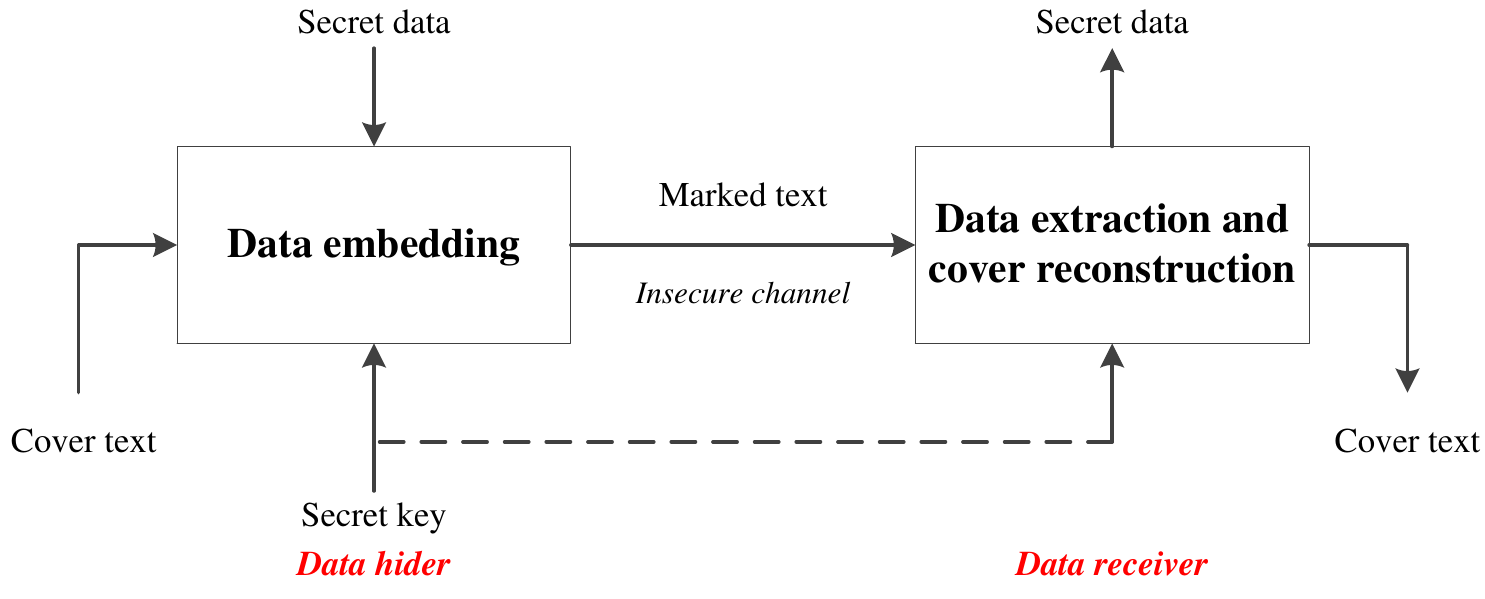}
\caption{Sketch for reversible data hiding in texts.}
\end{center}
\end{figure}

As shown in Fig. 1, RDH in texts can be briefly described as follows. The data hider needs to design such a data embedding algorithm that it accepts a given text (i.e., \emph{cover text}), a secret key and the secret data as input and thereafter outputs a new text carrying the secret data (i.e., \emph{marked text}). The resulting marked text will be sent to the data receiver via an insecure channel. According to the secret key, the data receiver is able to extract the secret data from the marked text and reconstruct the cover text without any error. In this way, RDH in texts is realized. One of the most important requirements for RDH in texts is that the marked text should be seemingly normal, i.e., the marked text should not introduce suspicious traces that will reveal the existence of secret information. It requires that the fluency and the semantic quality of the marked text should be satisfactory. Moreover, the marked text should be able to resist against statistical detection tools. 

A simple method of realizing RDH in texts is to change the text format in a lossless fashion. For instance, the white space between two adjacent words (or adjacent lines) can be adjusted to accommodate secret bits \cite{RDH:Text:Space, RDH:Text:Space2, RDH:Text:Space3}. By extracting secret bits from the marked text format, the original text can be further recovered since the words of the original text are unchanged. Another simple idea is to alter the characters within words of the cover text, which enables secret bits to be embedded but may result in incorrect words. In order to recover the cover text, these incorrect words caused by data embedding should be corrected. Nevertheless, abnormal text format and incorrect words will easily arouse suspicion from the adversary, thereby impairing the imperceptibility of secret information.

One could extend reversible embedding strategies originally designed for digital images to texts. For example, Liu \emph{et al.} \cite{RDH:IT} convert some words in the cover text into integers and then embed secret bits into the words by applying integer transform and difference expansion widely used in RDH in images. Even though this method ensures reversibility, the pure payload size is quite low due to the underflow/overflow problem. Moreover, though it can be improved so as to increase the payload size to a certain extent, new problems such as large side information will arise \cite{RDH:IT:Improved, Xiang:RDH:Text}. This indicates that it is not easy to extend RDH methods applied to digital images to texts.

With the fast development of deep learning \cite{DLNature} and natural language processing, mainstream DH methods applied to texts use a trained language model for facilitating embedding \cite{Kang:paper, Guo:paper, Yibiao:paper2, tianyu:arxiv, Zheng:SCN}. However, these methods cannot ensure reversibility. Just in recent, Chang \cite{Chang:2022} proposes a method to reversibly embed secret bits into a cover text by exploiting index expansion. In the method, a trained language model is used to predict the word of a position to be embedded, which allows the data hider to collect a list of prediction probabilities for all the candidate words. These words can be associated with an index according to the prediction probabilities. Thus, by determining the index matching the present bit to be embedded, the corresponding word can be selected as the output. Though the method shows good trade-off between data-embedding capacity and semantic distortion, it requires the data hider and the data receiver to share the pre-trained language model in advance, which may be not suitable for applications. On the one hand, a language model contains a number of parameters, meaning that the side information shared between the data hider and the data receiver is too much, which is not desirable in practice. On the other hand, the data hider may not want to share the trained language model with the data receiver since the trained language model can be treated as a private assert of the data hider. In addition, using a language model for data extraction corresponds to a high computational complexity, which is not good for practice.

Therefore, we urgently need to develop more efficient RDH schemes for texts. This has motivated the authors in this paper to propose a general framework for text based RDH. The main idea of the proposed framework is to generate such a marked text that the cover text can be reconstructed by collecting the words of some specific positions, and the words of the other positions can be processed to extract the secret information. As a result, the secret information and the original cover text can be successfully concealed within the marked text, and perfectly extracted from the marked text as well. Experimental results show that the proposed work ensures reversibility and provides good text quality and security, demonstrating the superiority. 

The rest of this paper is organized as follows. In Section II, we introduce the proposed framework. Then, in Section III, experimental results and analysis are provided to demonstrate the superiority and applicability of the proposed framework. Finally, we provide conclusion and discussion in Section IV. 

\begin{figure}[!t]
\begin{center}
\includegraphics[width=\linewidth]{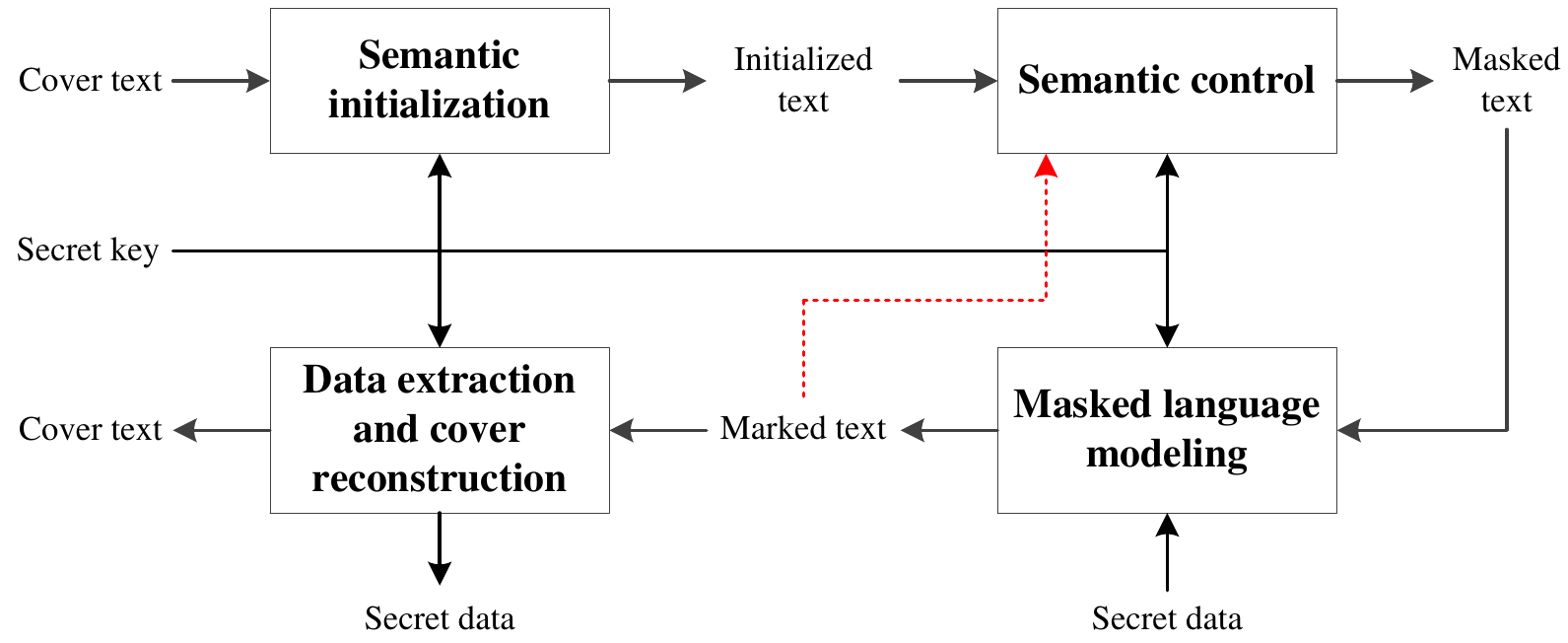}
\caption{General framework for the proposed method.}
\end{center}
\end{figure}

\section{Proposed Method}
The proposed method follows the framework shown in Fig. 1. However, different from mainstream methods that modify a given cover text to embed secret information, the proposed method uses the cover text as a ``semantic control key'' to guide us to generate a marked text carrying the secret information. Fig. 2 shows the general framework of the proposed method. It can be inferred from Fig. 2 that there are five important parts in the proposed framework, i.e., semantic initialization, semantic control, masked language modeling, data extraction and cover reconstruction. We detail each of them in the following. 

\subsection{Semantic Initialization}
Given a cover text and a secret (position) key, the purpose of semantic initialization is to produce an \emph{initialized text} that will be used for the subsequent data embedding procedure. In detail, let $\textbf{c} = \{c_1, c_2, ..., c_n\}$ denote the cover text, where $c_i$, $1\leq i\leq n$, is the $i$-th word (token) sampled from a very large vocabulary $V$. The secret position key $\textbf{p} = \{p_1, p_2, ..., p_n\}$ is an integer sequence that satisfies $1\leq p_1< p_2< ...< p_n$. The goal of semantic initialization is to generate the initialized text $\textbf{u} = \{u_1, u_2, ..., u_m\}$ that satisfies $p_n \leq m$ and 
\begin{equation}
u_i = 
\left\{\begin{matrix}
c_j &  \text{if}\ i = p_j \in \textbf{p},\\ 
\text{[MASK]} &  \text{otherwise}, 
\end{matrix}\right.
\end{equation}
where `[MASK]' denotes a special token. For example, assuming that \textbf{c} = \{I, do, .\}, \textbf{p} = \{1, 7, 12\} and $m$ = 12, \textbf{u} can be determined as \{I, [MASK], [MASK], [MASK], [MASK], [MASK], do, [MASK], [MASK], [MASK], [MASK], .\}.

\subsection{Semantic Control}
The purpose of semantic control is to generate a \emph{masked text} that will be fed into the subsequent masked language modeling module. It is free to design the algorithm of semantic control, namely, there is no strict restriction on the masked text except that the words corresponding to the cover text should not be changed. The proposed method simply sets the masked text as the initialized text or the \emph{temporary marked text}, which is due to the reason that the subsequent embedding method is an iterative process (refer to the next subsection).

\subsection{Masked Language Modeling (Data Embedding)}
The purpose of masked language modeling is to generate a marked text carrying secret information, which is an iterative process. In each iteration, we use a pre-trained language model to generate a temporary marked text based on the masked text and secret data to be embedded. The difference between the masked text and the temporary marked text is that exactly one masked position of the masked text is replaced by a word. The temporary marked text will be used to replace the masked text for the next iteration unless all masked positions are processed. It is noted that masked language modeling is a popular strategy for text generation in mainstream natural language processing tasks. However, in this paper, the masked language modeling strategy is used for generating texts carrying additional information. That is why the masked language modeling procedure in this paper is actually equivalent to data embedding.

The pseudo-code of data embedding is shown in Algorithm 1. It can be seen that by predicting the word of each masked position, we can generate the final marked text. However, as shown in Line 5 of Algorithm 1, we should use an information encoding method to determine the word of a specific masked position, which is free for us to design. The goal of information encoding is to select a candidate word from a list of candidate words to replace the special token `[MASK]' and meanwhile match the secret data to be embedded. Since it is not the main interest of this paper, we use the existing information encoding method for simplicity. In experiments, we will evaluate various information encoding strategies.

\subsection{Data Extraction}
The goal of data extraction is to extract the secret information \textbf{b} from the marked text. It depends on the information encoding method shown in Line 5 of Algorithm 1. Specifically, if the mapping relationship between words and secret bits is based on the prediction probabilities of candidate words, both information encoding and information decoding are controlled by the pre-trained language model $\mathcal{M}$. It indicates that the pre-trained language model $\mathcal{M}$ should be shared between the data hider and the data receiver because the data receiver should use the language model to reconstruct the mapping relationship between words and secret bits. Otherwise, there is no need for the data hider to share $\mathcal{M}$ with the data receiver. Obviously, from the viewpoint of practical use, it is more desirable that only the data hider holds the language model. However, from the viewpoint of embedding performance, sharing the language model may be better. In experiments, we find that the former already achieves a competitive performance, implying that not sharing the language model is a good strategy for RDH.

\subsection{Cover Reconstruction}
Reconstructing the original cover text is straightforward. It can be described as
\begin{equation}
c_i = s_{p_i}, \forall 1\leq i\leq n,
\end{equation}
where the position key $\textbf{p} = \{p_1, p_2, ..., p_n\}$ should be shared between the data hider and the data receiver in advance.

\begin{algorithm}[!t]
 \caption{Pseudocode for the data embedding procedure}
 \begin{algorithmic}[1]
	\renewcommand{\algorithmicrequire}{\textbf{Input:}}
	\renewcommand{\algorithmicensure}{\textbf{Output:}}
	\REQUIRE Initialized text $\textbf{u}$, language model $\mathcal{M}$, secret bits $\textbf{b}$
	\ENSURE Marked text $\textbf{s}$.
	\STATE Initialize \textbf{s} = \textbf{u}
    \FOR{$i=1,2,...,m$}
        \IF {$s_i$ = [MASK]}
            \STATE Apply $\mathcal{M}$ and \textbf{s} to generate a list of candidate words each associated with a prediction probability for $s_i$ // \emph{notice that \textbf{s} can be called as a masked text here}
            \STATE Determine a candidate word $w$ carrying a prefix of \textbf{b} according to an information encoding method
    	    \STATE Update $\textbf{s}$ by applying $s_i = w$ // \emph{notice that \textbf{s} can be called as a temporary marked text here}
    	    \STATE Remove the embedded prefix from $\textbf{b}$
    	\ENDIF
    \ENDFOR
	\RETURN $\textbf{s}$
 \end{algorithmic}
\end{algorithm}

\section{Experimental Results and Analysis}
In this section, we are to provide experimental results and analysis for performance evaluation. 

\subsection{Setup and Evaluation Metrics}
We use Python and PyTorch for simulation. For the language model, we use Google's $\text{BERT}_\text{Base, Uncased}$ model and Hugging Face's transformers package \cite{transformer:Paper} based on the default settings. The benchmark dataset BookCorpus \cite{dataset}, a large text corpus widely used in natural language processing tasks, consisting of around 18,000 books with rich topics and different fine-grained semantic information, is used for experiments. We randomly select 10,000 cover texts for RDH. One thing to note is that the length of each cover text, i.e., $n$, needs to be controlled within a reasonable range, which can reduce the computational complexity while ensuring the quality of the marked text. And, for simplicity, we assume that the length of the marked text is a multiple of that of the cover text, namely, $n|m$. $m$ may be adaptively increased to become a larger integer not divided by $n$ so that the marked text ends with a stop token.

To evaluate the quality of the marked texts generated by the proposed method, the standard measure of perplexity (PPL) is used. The PPL calculates the average log probability of each word in the sentence. Generally, the lower the perplexity is, the more natural the generated text is and the more secure the generated text will be. We use GPT-2 \cite{GPT:paper} to determine the average PPL of all marked texts as the judging criterion. In order to evaluate the security, we use 10,000 natural texts and 10,000 marked texts with the same length range and similar grammatical structure for steganalysis with the method in \cite{RNN:steganalysis}. Two common metrics: accuracy (Acc) and F1 score (F1) \cite{GNN:steganalysis, LM:steganalysis}, are used to measure the ability to resist steganalysis tools. Yet another indicator is the data-embedding payload, which is defined as the average bits carried by each word (i.e., \emph{bits per word, bpw}) and is expected to as high as possible. 

\subsection{Information Encoding Strategies}
For fair comparison, we use different information encoding strategies for simulation. Four representative information encoding strategies, i.e., block coding \cite{block:coding}, Huffman coding \cite{Zheng:SCN}, adaptive dynamic grouping (ADG) \cite{Zhang:ADG}, and bins coding \cite{Bins:paper} are tested in our experiments. For self-contained, we briefly describe the technical details of them in the following.

\subsubsection{Block Coding}
Given a set of candidate words, the block coding method assigns $k$-bit binary codes from 0 to $2^k-1$ to $2^k-1$ words selected from the set according to the prediction probabilities. For example, for a specific masked position, by sorting all the candidate words in the vocabulary $V$ according to their prediction probabilities in a descending order, each of the top-$2^k$ words can be associated with a binary code with a length of $k$. We need to use a threshold $t_p$ to control the number of usable words in $V$, i.e., for a given masked position, only those words in $V$ with a prediction probability larger than $t_p$ can be used for carrying secret bits. On the other hand, the data receiver should hold $t_p$ and the (masked) language model so that he can extract the embedded data from the marked text. 

\subsubsection{Huffman Coding}
Huffman coding is a consistency coding technique. While the above block coding method assigns fixed-length codes to the words, Huffman coding assigns codes with an indefinite length to words according to the prediction probabilities. Same as block coding, Huffman coding needs $t_p$ and the language model to embed data and extract data.

\subsubsection{Adaptive Dynamic Grouping (ADG)}
The target of ADG is to group all words (tokens) in the vocabulary into a certain number of groups so that each group represents a unique secret binary stream. To embed secret data, for a masked position, a word is sampled from the corresponding group as the output. For data extraction, the data receiver should have the language model and vocabulary, and know the grouping algorithm. For consistency, $t_p$ is used to control the number of usable words. 

\subsubsection{Bins Coding}
Bins coding maps all words in the vocabulary into a binary stream in advance. During data embedding, for a masked position, among a list of words that match the secret data to be embedded, the one with the largest prediction probability is used as the output. For example, the vocabulary $V$ can be divided into two disjoint subsets $V_0$ and $V_1$, where $|V_0| \approx |V_1|$. The words in $V_b$ are mapped to the secret bit $b\in \{0, 1\}$. During data embedding, if the secret bit is $b$, the word with the largest prediction probability in $V_b$ is selected as the output. Obviously, it is easy for the data receiver to extract the secret data without knowing the prediction probability of the word. In other words, compared with the above methods, there is no need for the data hider and the data receiver to share the language model, which significantly reduces the side information. By default, in our experiments, we divide $V$ into two subsets, implying that, we use each word to carry one bit. It is pointed that there is no need for the data hider and the data receiver to 
store $V_b$ because the mapping relationship between words and secret bits can be realized by applying a hash function, which is very convenient for data extraction. 

\begin{table}[!t]
\renewcommand{\arraystretch}{1}
\centering
\caption{Some examples for the marked text, where $m/n = 4$.}
\begin{tabular}{c|c|c}
\hline\hline
Encoding & $t_p$ & Marked text \\
\hline
\multicolumn{1}{c|}{\multirow{3}{*}{Block}} &
$0.02$  & \emph{\textbf{I} was going to try to \textbf{do} the same for myself \textbf{.}} \\
&$0.03$  & \emph{\textbf{I} have no way to really \textbf{do} the same without him \textbf{.}} \\
&$0.04$  & \emph{\textbf{I} do this , but I do not \textbf{do} this now \textbf{.}} \\
\hline
\multicolumn{1}{c|}{\multirow{3}{*}{Huffman}} & 
$0.02$  & \emph{\textbf{I} know what the police will \textbf{do} if he comes here \textbf{.}}\\
& $0.03$  & \emph{\textbf{I} do it . you always \textbf{do} it . I know \textbf{.}}\\
& $0.04$  & \emph{\textbf{I} always \textbf{do} it for a living \textbf{.}}\\
\hline
\multicolumn{1}{c|}{\multirow{3}{*}{ADG}} &
$0.02$  & \emph{\textbf{I} have no one willing to \textbf{do} a little crazy thing \textbf{.}} \\
&$0.03$  & \emph{\textbf{I} did the same thing you \textbf{do} for the same reason \textbf{.}} \\
&$0.04$  & \emph{\textbf{I} did . but you can \textbf{do} that for her too \textbf{.}} \\
\hline
Bins & - &	\emph{\textbf{I} know the way they can \textbf{do} it to me now \textbf{.}}  \\
\hline
Top-1 & - &	\emph{\textbf{I} know what I have to \textbf{do} to keep her safe \textbf{.}} \\
\hline\hline
\end{tabular}
\end{table}

\begin{table}[!t]
\renewcommand{\arraystretch}{1}
\centering
\caption{Mean payload sizes (bpw) due to different information encoding strategies, where $n\in [4, 8]$.}
\begin{tabular}{c|c|c|c|c}
\hline\hline
Encoding strategy & $t_p$ & $m/n = 3$& $m/n = 4$ & $m/n = 5$ \\
\hline
\multicolumn{1}{c|}{\multirow{3}{*}{Block}} &
$0.02$  & 1.2113 & 1.3822 & 1.4691 \\
&$0.03$  & 0.9538 & 1.0794 & 1.1402 \\
&$0.04$  & 0.7865 & 0.8839 & 0.9315 \\
\hline
\multicolumn{1}{c|}{\multirow{3}{*}{Huffman}} & 
$0.02$  & 1.2269 & 1.3946 & 1.4792\\
& $0.03$  & 1.0166 & 1.1464 & 1.2154\\
& $0.04$  & 0.8576 & 0.9620 & 1.0168\\
\hline
\multicolumn{1}{c|}{\multirow{3}{*}{ADG}} &
$0.02$  & 0.4734 & 0.5335 & 0.5507  \\
&$0.03$  & 0.3561 & 0.3977 & 0.4113  \\
&$0.04$  & 0.2742 & 0.3015 & 0.3104 \\
\hline
Bins & - &	0.6507 & 0.7259 & 0.7660  \\
\hline\hline
\end{tabular}
\end{table}

\begin{table}[!t]
\renewcommand{\arraystretch}{1.2}
\centering
\caption{Mean PPLs due to different information encoding strategies, where $n\in [4, 8]$.}
\begin{tabular}{c|c|c|c|c}
\hline\hline
Encoding strategy & $t_p$ & $m/n = 3$ & $m/n = 4$ & $m/n = 5$ \\
\hline
\multicolumn{1}{c|}{\multirow{3}{*}{Block}} & 
$0.02$  & 223.8311 	& 157.5531 	& 128.8145  \\
& $0.03$  & 202.9565 & 134.4043 & 103.8719  \\
& $0.04$  & 187.6439 & 118.5401	& 91.9489 \\
\hline
\multicolumn{1}{c|}{\multirow{3}{*}{Huffman}} & 
$0.02$  & 202.6945 & 138.0253 & 108.4690 \\
& $0.03$  & 189.8209 & 126.0925 & 97.6912 \\
& $0.04$  & 180.3138 & 116.3862 & 89.3254\\
\hline
\multicolumn{1}{c|}{\multirow{3}{*}{ADG}} &
$0.02$  & 260.7759 & 189.2484 & 157.2128   \\
&$0.03$  & 216.1935 & 153.3190 & 125.8440   \\
&$0.04$  & 199.3847	& 134.2199 & 104.6727 \\
\hline
Bins & - & 274.5535 & 179.6919 & 134.2634   \\
\hline\hline
\end{tabular}
\end{table}

\begin{table*}[!t]
\renewcommand{\arraystretch}{1}
\centering
\caption{Detection accuracy due to different information encoding strategies, where $n\in [4, 8]$.}
\begin{tabular}{c|c|c|c|c|c|c|c}
\hline\hline
\multicolumn{1}{c|}{\multirow{2}{*}{Encoding strategy}} & 
\multicolumn{1}{c|}{$m/n$} &
\multicolumn{2}{c|}{$3$} & \multicolumn{2}{c|}{$4$} & \multicolumn{2}{c}{$5$}\\
\cline{2-8}
& $t_p$ & Acc  &  F1 & Acc  &  F1 & Acc  &  F1  \\
\hline
\multicolumn{1}{c|}{\multirow{3}{*}{Block}} &
$0.02$  & 0.8917 & 0.8934 & 0.9213 & 0.9211 & 0.9418 & 0.9416\\
& $0.03$ & 0.9040 & 0.9030 & 0.9247 & 0.9254 & 0.9473 & 0.9469\\
& $0.04$ & 0.8938 & 0.8925 & 0.9232 & 0.9216 & 0.9425 & 0.9438\\
\hline
\multicolumn{1}{c|}{\multirow{3}{*}{Huffman}} &
$0.02$  & 0.8992 & 0.8987 & 0.9195 & 0.9202 & 0.9435 & 0.9441\\
& $0.03$  & 0.9020 & 0.9028 & 0.9250 & 0.9260 & 0.9535 & 0.9540 \\
& $0.04$ & 0.9002 & 0.9018 & 0.9248 & 0.9245 & 0.9497 & 0.9495\\
\hline
\multicolumn{1}{c|}{\multirow{3}{*}{ADG}} &
$0.02$  & 0.9255 & 0.9267 & 0.9475 & 0.9478 & 0.9537 & 0.9542\\
& $0.03$  & 0.9127 & 0.9143 & 0.9397 & 0.9407 & 0.9540 & 0.9538\\
& $0.04$ & 0.9087 & 0.9069 & 0.9398 & 0.9402 & 0.9573 & 0.9568\\
\hline
Bins & - & 0.9098 & 0.9094 & 0.9305 & 0.9290 & 0.9515 & 0.9521\\
\hline\hline
\end{tabular}
\end{table*}

\subsection{Results and Analysis}
We first provide some examples to verify the feasibility of the proposed method. The examples are in Table I, where we have \textbf{c} = \{I, do, .\}, \textbf{p} = \{1, 7, 12\}, $m = 4n = 12$. Regardless of the length of the embedded payload, it can be inferred that the generated marked texts have satisfactory quality. In Table I, ``Block'', ``Huffman'', ``ADG'' and ``Bins'' are corresponding to block coding, Huffman coding, adaptive dynamic grouping and bins coding, respectively. ``Top-1'' means that no information is embedded, i.e., each masked position was always filled with the word having the largest prediction probability. 

To measure the embedding payload, we empirically limit $n$ to range $[4, 8]$ for simplicity. Table II shows the mean payload sizes due to different information encoding strategies, from which we can find that different information encoding strategies result in different payload sizes. It is reasonable because different strategies have different information utilization. As shown in Table II, the size of the mean payload will decline when $t_p$ increases. The reason is that a higher $t_p$ indicates that less words are used to carry secret data, thereby resulting in a lower payload size. For bins coding, though the payload size is less than that of block coding and Huffman coding in most cases, the payload size can be increased by dividing the vocabulary $V$ into more disjoint subsets. For example, by dividing $V$ into four disjoint subsets, each word can be used to carry two bits, resulting in a double payload size, thereby outperforming other methods in terms of embedding payload. In addition, a higher $m/n$ means that more masked positions are used for data embedding, accordingly resulting in a higher payload size, which has been verified by Table II.

The mean PPLs of the marked texts due to different information encoding strategies are also determined. As shown in Table III, the PPL value declines when $t_p$ increases, which is due to the reason that a higher $t_p$ is corresponding to a lower embedding payload size, resulting in a lower distortion to the generated marked text. Moreover, the PPL gradually declines when $m/n$ increases, which is due to the reason that a longer text tends to become more fluent because of more contexts.

In order to further measure the security of the marked texts, we evaluate the ability of the marked text to resist steganalysis. 10,000 natural texts and 10,000 marked texts are selected to form the sample set, divided into training set, validation set and testing set according to the ratio of 6:1:3, evaluated on the testing set by the model with the highest validation accuracy. The results are provided in Table IV. It can be inferred that the detection accuracy and F1 score gradually decline in most cases when $t_p$ increases. It is reasonable because a higher $t_p$ corresponds to a lower embedding payload size, resulting in a lower statistical difference between natural texts and marked texts, making the detection accuracy and F1 score lower. The detection accuracy and F1 score gradually increase when $m/n$ increases, which is due to the reason that longer texts are more likely to expose statistical anomalies. However, though different information encoding strategies result in different steganalysis performance, the performance difference between different strategies are close to each other in most cases. From the viewpoint of practical use, it will be more desirable to use information encoding strategies that does not require the data hider and the data receiver to share much side information. In other words, in terms of practicality, based on Table IV, it is more desirable to use bins coding for RDH. 

\section{Conclusion}
In this paper, we propose a novel framework for reversible data hiding in texts, which is totally different from previous methods that extend reversible embedding strategies designed for digital images to texts. In the proposed framework, by distributing the cover text into the masked text according to the position key, the masked positions of the masked text are filled with marked words to embed secret data. Experimental results show that secret data can be extracted from the marked text without any error. The original cover text can be perfectly recovered with the pre-shared position key. By fine-tuning the parameters, a sufficient payload can be achieved. Furthermore, the generated marked texts have good quality and satisfactory ability to resist against steganalysis. In future, we will further improve the embedding performance. We hope this framework could inspire more advanced works. 

\section*{Acknowledgement}
This work was financially supported in part by the Opening Project of Guangdong Province Key Laboratory of Information Security Technology under Grant number 2020B1212060078. It was also supported in part by the National Natural Science Foundation of China under Grant number 61902235, and the Shanghai Chenguang Program under Grant number 19CG46.

\end{document}